\documentclass[]{aa}  
\usepackage{graphicx,epsfig}
\usepackage{txfonts}
\usepackage{natbib}
\usepackage{longtable}
\bibpunct{(}{)}{;}{a}{}{,} 
\begin{document}
  \title{The nature of V39: an LBV candidate or LBV impostor in the very low metallicity galaxy IC\,1613? \thanks{Based on observations obtained at 
  the ESO VLT for Programmes 078.D-0767 and 080.D-0423.}}

   \author{A. Herrero \inst{1,2}
          \and 
 	    M. Garcia \inst{1,2}
           \and 
  	    K. Uytterhoeven \inst{1,3} 
          \and 	    
          F. Najarro \inst{4} 
          \and 
	   D.J. Lennon\inst{5}
	   \and
	   J.S. Vink\inst{6}
	   \and
	   N. Castro\inst{1,2}
	  }

   \offprints{A. Herrero (\email{ahd@iac.es})}

  \institute{Instituto de Astrof\'{i}sica de Canarias, C/ V\'{i}a L\'{a}ctea s/n, E-38200 La Laguna, Tenerife, Spain.
         \and
	 	Departamento de Astrof\'{i}sica, Universidad de La Laguna, Avda. Astrof\'{i}sico
		Francisco S\'anchez s/n, E-38071 La Laguna, Tenerife, Spain.
	\and
		Laboratoire AIM, CEA/DSM-CNRS-Universit\'e Paris Diderot; CEA, IRFU,
SAp, centre de Saclay, F-91191, Gif-sur-Yvette, France
	 \and
	 	Centro de Astrobiolog\'{i}a (CSIC-INTA), Ctra. de Torrej\'on a Ajalvir km-4, E-28850, Torrej\'on de Ardoz, Madrid, Spain 
	\and
		 ESA, Space Telescope Science Institute, 3700 San Martin Drive, Baltimore, MD 21218, USA
	\and
		Armagh Observatory, College Hill, Armagh BT61 9DG, Northern Ireland
    }

   \date{}

 
  \abstract
{Very few examples of luminous blue variable (LBV) stars or LBV candidates (LBVc) are known, particularly at metallicities below
the SMC. The LBV phase is crucial for the evolution of massive stars, and its behavior with metallicity is poorly known. V39 in 
IC 1613 is a well-known photometric variable, with B-band changes larger than 1mag. over its period. The star, previously 
proposed to be a projection of a Galactic W Virginis and an IC 1613 red supergiant, shows features that render it a possible LBVc.}
{We aim to explore the nature of V39 and estimate its physical parameters.}
{We investigate mid-resolution blue and red VLT-VIMOS spectra of V39, covering a time span of 40 days, and perform a quantitative 
analysis of the combined spectrum using the model atmosphere code CMFGEN.}
{We identify strong Balmer and \ion{Fe}{ii} P-Cygni profiles, and a hybrid spectrum resembling a B-A 
supergiant in the blue and a G-star in the red. No significant Vrad variations are detected, and the spectral changes are small over the photometric period. 
Our analysis places V39 in the low-luminosity part of the LBV and LBVc region, but it is also consistent with a sgB[e] star. 
From this analysis and the data in the literature we
find evidence that the [$\alpha$/Fe] ratio in IC 1613 is slightly lower than solar.}
{The radial velocity indicates that V39 belongs to IC 1613. The lack of Vrad changes 
and spectroscopic variations excludes binary scenarios. 
The features observed are not consistent with a W Virginis star, 
and this possibility is also discarded. We propose that the star is a 
B-A LBVc or sgB[e] star surrounded by a thick disk precessing around it.
If confirmed, V39 would be the lowest metallicity resolved LBV candidate known to date. 
Alternatively, it could represent a new transient phase of massive star evolution, an LBV impostor.}

  
	\authorrunning{A. Herrero et al.}
	\titlerunning{The nature of V39}
   \maketitle


\section{Introduction}

Luminous blue variables (LBVs) constitute a short and rare phase in the life of massive stars, in which the stars suffer great mass loss with 
unpredictable outbursts and mass eruptions. Apart from the stages in which a nuclear runaway is produced in the stellar core, LBVs seem to be the shortest phase in the massive star's evolution, even shorter than Wolf-Rayet phases, with estimated lifetimes of 10$^4$-10$^5$ years. 
In spite of its brevity, this phase is extremely important for the evolution of the star and its surroundings, because it returns important amounts of
processed material to the ISM, often forming a shell or a nebula around the star. 
Moreover, the loss
of matter modifies the star's subsequent evolution and it has been very recently suggested that LBVs might be {\it direct} SNe progenitors
(see \cite{kotak06}, \cite{vink09}), which may have an impact on our current ideas of massive star and galaxy evolution in the Universe. 

It is a tautology to state that LBVs are luminous (log(L/L$_\odot$ $\geq$ 5.4), blue ((B-V)$_0$ $\leq$ 0.1) and variable stars. 
Nevertheless, their variations can be considerably different in frequency and intensity, going from intense 
and sudden eruptions in which the stellar magnitude increases by two magnitudes or more (like $\eta$ Car) to cyclic variations of 1-2 magnitudes
in time scales of years to decades (like the prototype S-Dor).  These photometric changes are accompanied by the corresponding spectroscopic
ones, in which the effective temperature varies and the bolometric luminosity may or may not remain constant \citep{vink09}. 
However, outbursts and eruptions are not very common, and sometimes the star remains 
stable for centuries, as it is the case for P-Cygni. Thus, rather than from photometric variations, they are often searched for by means of their 
P-Cygni profiles in the Balmer lines (which are a signature of the presence of strong mass-loss processes), emission line profiles (at lower
resolutions or NIR wavelengths) and of 
other lines which depend on the temperature of the star. When the star shows a characteristic spectrum with P-Cygni profiles, but no relatively large 
photometric and spectroscopic variations (or we do not have information about such variations), we speak of an LBV candidate, or LBVc. 
Sometimes, other signatures (like IR excess or nebulosity, see f.e. \cite{clark05}) are considered further signs for the classification as LBVc.

The mechanism producing LBVs is not known, neither is the exact duration of the phase (and hence, the exact impact on the star's life). 
It is assumed that the stellar mass and, very especially, the mass-loss rate play a role. As the winds of massive stars are driven by the momentum gained 
by the atoms from the radiation field through the absorption of photons by metal lines, LBVs should be rare at low metallicities.  
This is consistent with no LBVs  or LBV candidates known 
in the Local Group at metallicities below that of the Small Magellanic Cloud (SMC). 
While this fact could be at least partly the result of the low number of massive stars in
such galaxies, the sheer presence of an LBV at low Z should be considered significant, as
their presence offers clues for mass-loss processes that could have been relevant in the early Universe.

Beyond the Local Group few relevant examples of LBVs or possible LBVs are known. \citet{drissen01} (see also \citet{petit06}) have analyzed
the star V1 in NGC 2363, a galaxy at about 3.3 Mpc (\citet{tolstoy95}, \citet{kara02}) with 12+log(O/H)= 7.9 \citep{gonzalez94} 
or Z= 0.25 Z$_\odot$ \citep{luridiana99}. HST spectra were used for the analysis to avoid contamination because the star is in a dense 
stellar environment with strong nebulosity. NGC 2363-V1 shows strong photometric and spectroscopic variations \citep{petit06}, thus fulfilling all 
requirements to be called an LBV. Using the CMFGEN code \citep{hillier98}, \citet{drissen01} obtained  T$_{\rm eff}$= 11000 -- 13000 K,
log(L/L$_\odot$)= 6.35 -- 6-40 and v$_\infty$= 325 -- 290 km s$^{-1}$ for spectra collected in Nov. 1997 and July 1999, respectively,
fully in agreement with the values expected for an LBV in transition. From the iron spectrum they get Z$\approx$ 0.2 Z$_\odot$, intermediate between
\citet{gonzalez94} and \citet{luridiana99}.

Very interesting, but
less evident is the case of two very recent objects studied in very metal poor galaxies: knot 3 in DDO 68 (\citet{pustilnik08}; \citet{izotov09}) and
PHL 293B \citep{izotov09}. DDO 68 is a blue compact dwarf (BCD) galaxy at a distance of at least 5.9 Mpc (\citet{makarova98} determined its
distance from the average magnitude of the three brightest blue stars). 
Knot 3 (DDO 68-3)  is one of its strong HII regions, for which \citet{pustilnik08} give 12+log(O/H)= 7.10$\pm$0.06 and 
\citet{izotov09} 7.15$\pm$0.04.  In spectra of this knot separated by about three years, 
\citet{pustilnik08} detected spectral variations (broader emission in the Balmer lines and \ion{He}{i} 
emission not previously present) at resolutions of 6 and 12 \AA~ FWHM 
that they attributed to the outburst of an LBV inside the region. This explanation was adopted and further investigated by \citet{izotov09}, who 
estimated the absolute magnitude, H$_\alpha$ luminosity and wind terminal velocity from APO and MMT spectra at resolutions of 7 and 3 \AA~ FWHM.
However, note that the 1.5" slit width used by \citet{izotov09} corresponds to about 45 pc at the distance of DDO 68 and therefore the
observed spectrum is a composite of the HII region and the underlying stellar population (including the possible LBV). 

A similar problem is found
in the second object presented by \citet{izotov09}, PHL 293B. This is another low-metallicity BCD (12+log(O/H)= 7.66$\pm$0.04 according to
\citet{izotov07} and 7.72$\pm$0.01 according to \cite{izotov09}) but this time located at 21.4 Mpc (NED database). The spectra
of \cite{izotov09} were obtained with a slit width of 1.0", which corresponds to more than 100 pc. The oxygen abundance quoted by \cite{izotov09}
has been obtained from a high resolution (0.2 \AA~ FWHM) VLT-UVES spectrum that
also displays broad emissions at the positions of H$_\alpha$, H$_\beta$ and H$_\gamma$ and blue-shifted absorptions in all Balmer lines. No He broad
emissions are seen. Neither the DDO 68-3 nor the PHL 293B spectra show metal lines that could be attributed to a star. In both cases, the
wind terminal velocities quoted by \cite{izotov09} ($\approx$ 700-850 km s$^{-1}$) are significantly larger than the usual range of LBVs ( 100-250 km$^{-1}$;
500 km s$^{-1}$ for $\eta$ Car), which is attributed by the authors to the very low metallicity. 

The mere presence of LBVs at these low metallicities represents a challenge for the theory, as there are few metals to drive the wind. 
Other processes have been suggested to cause the LBV outburst, like the metallicity-independent continuum-driven (instead of line-driven) wind 
proposed by \citet{owocki04} (see also \cite{smith06}). Note however that the star needs to be close to the Eddington limit for
the mechanism to operate. There is a clear interest in investigating these processes, confront them with
observations and apply the results to early Universe and Population III objects. Obviously, it would be very interesting to obtain spatially resolved spectra of the
stars in DDO 68-3 and PHL 293B, both to confirm that the changes in the composite spectra are due to an
LBV and to study the evolutionary processes at very low metallicity. However, because of the large distances involved, we have to look closer
for resolved stars, such as NGC 2363-V1.
  
IC\,1613 is a dwarf irregular galaxy in the Local Group  with a distance modulus of (m-M)$_0$= 24.27 
\citep{dolphin01}. It has a metallicity of log(O/H)+12=  7.80$\pm$0.10 as determined from its B-supergiants \citep{bresolin07}, 
whereas nebular studies vary between 7.60 and 7.90 \citep{lee03}. It is therefore clearly
below the SMC metallicity and intermediate between that of NGC 2363 and PHL 293B.  Therefore, the observation of a resolved LBV in such a 
poor-metal galaxy would constitute a step forward, adding a second object to the list of LBVs at metallicities below that of the SMC 
and lowering the present metallicity limit for resolved LBVs.

IC\,1613 shows a recent and intense burst of massive star
formation, particularly in its NE part. We have recently published a new catalog of OB associations in IC\,1613 \citep{garcia09} 
and their physical properties (Garcia et al., in prep.) as part of
our effort to carry out an in-depth study of the young population of IC\,1613. We also obtained spectra of some
stars in this galaxy. In the field of IC\,1613 we find the well-known variable star V39. 
In Table~\ref{stardata} we offer an overview of its photometric data.

\begin{table*}[htdp]
\caption{Photometric data for V39. \citet{udalski01}, \citet{pietr06} and \citet{garcia09} give mean values.}
\centering
\begin{tabular}{cccccccl}
\hline \hline
 U & B & V & R & I & J & K & Ref \\
 \hline
            & 18.6 - 19.9 &                   &           &                     &      &   & \cite{sandage71}\\
           &                   &  18.851      &            & 17.617         &   &   & \citet{udalski01}\\
           &                   & 18.5 - 19.0 &            & 17.45 - 17.7 &   &   & \citet{mantegazza02}\\
           &                   &                   &            &                     & 16.815 & 15.818 & \citet{pietr06}\\
19.43  & 19.62         & 19.00         &  18.37 & 17.75           &     &   & \citet{garcia09}\\
\hline
\end{tabular}
\label{stardata}
\end{table*}%
The nature of V39 has remained unclear since its discovery. \cite{sandage71}, 
in his analysis of IC1613 based on unpublished previous work by Baade, points out that 
V39 is the only peculiar variable in the field, because of its inverted $\beta$-Lyrae light curve. 
This was already noticed by Baade, who never considered it to be a Cepheid, not only because of this
peculiar light curve, but also because it was too bright to fit the P-L relation compared to other Cepheids 
with the same period. Consequently, in spite of a suggestion by \cite{sandage71} (that implied a
new P-L relation and a change in the distance modulus to bring it into agreement with other Cepheids)
most authors did not consider it a genuine Cepheid. \citet{udalski01} used its position
in the I vs (V-I) color-magnitude diagram as a new argument to discard it as a Cepheid. 
Very recently, \citet{pietr06} have also shown that V39 is too bright to fit the Cepheids P-L relation in the infrared.

\citet{antonello99} showed that the combined light curve from different observers is the result of the superpositon of two periods: a long
period of 1123 days and a short period of 28.699 days. They showed that by subtracting the long period from the
light curve, the remaining one did not show a clear difference between the primary and secondary maxima, so that the
short period could be halved. Based on these results, \cite{mantegazza02} proposed V39 to be 
actually the casual overlapping of a distant Galactic W Vir star and a
red supergiant belonging to IC\,1613. If confirmed, the W Vir star would be at a distance of at least 115 Kpc, being the most distant Galactic star from
the Galactic plane, and raising the question about the true extent of the Galactic halo. 

In this article we show that V39 does actually belong to IC\,1613 and analyze its possible nature, including the possibility 
that it is an LBV candidate, the first one known in IC\,1613.  If confirmed, it would be the (resolved) LBV candidate
within the lowest metallicity environment. The observations are presented in Sect.~\ref{obs}, a description of the spectrum is offered in Sect.~\ref{spec} and
the nature of V39 is then discussed in Sect~\ref{nature}. An estimation of the stellar parameters is offered in Sect.~\ref{param} and the
evolutionary status and other properties are discussed in Sect.~\ref{discus}. Finally, the conclusions are presented in Sect.~\ref{conc}.
  

\section{Observations and data reduction}
\label{obs}

Observations were performed with VIMOS at VLT in MOS (multi-object spectroscopy) mode.
The HR-Blue and HR-Orange gratings were used, resulting in a resolution of R$\sim 2050$
and R$\sim 2150$, respectively, and a spectral wavelength coverage of 3870 to 7240 \AA.   
In total $3\times19$ blue spectra were observed from October 5 to November 5, 2007,
and $2\times10$ red spectra from November 5 to November 13, 2007. Integration times
for individual exposures were 650 sec for the blue spectra and 1005 to 1110 sec for the
red spectra. The three (two) consecutive exposures in the blue (red) were coadded. The
coadded spectra were extracted with standard IRAF\footnote{IRAF is
distributed by the National Optical Astronomy Observatory, which is operated by the
Association of Universities for
Research in Astronomy, Inc., under cooperative agreement with  the National Science
Foundation.} procedures after wavelength
calibration and cosmic ray removal. Next, a barycentric correction and a correction for
the systemic velocity of IC1613 ($-$234$\pm$1 km s$^{-1}$, \cite{lu93}) were applied. Finally, all 19 (10) resulting spectra were
coadded to a single blue (red) spectrum. The resulting spectra, with an average SNR of 80, were then rectified.

\begin{figure*}
     \includegraphics[angle=0,width=14.0cm]{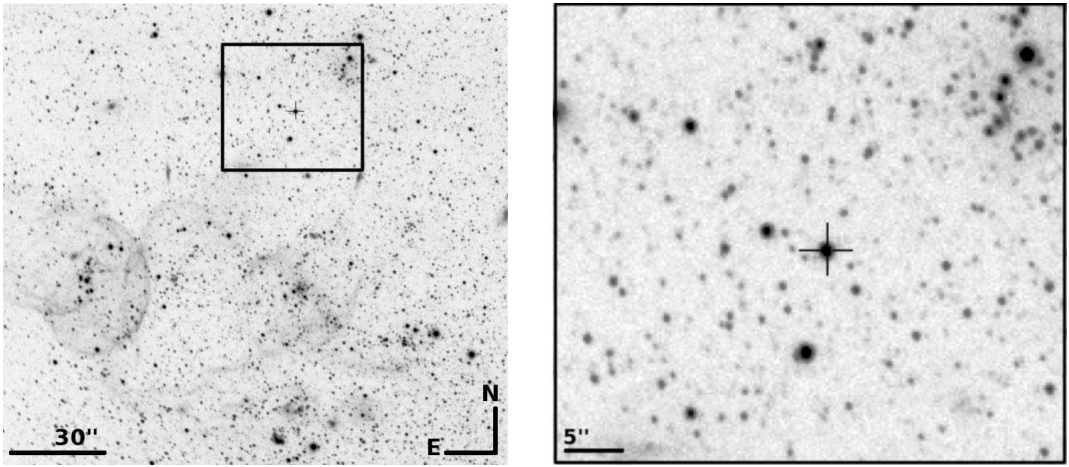}
     \caption{Finding chart for V39: R band VLT-VIMOS image of IC~1613, showing the active star-forming NE lobe of the galaxy (left)
and a zoom into V39's neighborhood (right), indicated by the square in the left panel. In both panels the position of V39 is marked with a cross. The slit for the observation 
was placed in the North-South direction; there is no resolved nearby star that could contaminate the spectrum in this direction.}
	\label{finding}
  \end{figure*}

Figure~\ref{finding} shows the finding chart of V39. The star is located at $\alpha$(2000)= 01h 05m 2.04s, $\delta$(2000)= 02$^{\deg}$ 10$^{\prime}$ 24.7$^{\prime\prime}$ (from the astrometry by \citet{garcia09}). It clearly lies in a zone of intense recent star formation, with strong
bubbles visible in Fig.~\ref{finding}. Table~\ref{stardata} gives the main photometric data of the star, collected from the literature.
Details for our astrometry and photometry are given in  \citet{garcia09}. At the time of the photometric observations
(July, 30th, 2006), the star was in a state of minimum brightness (see Sect.~\ref{nature}). 
Note that photometry and spectroscopy were obtained with a time difference of more than one year.

Unfortunately, no HST, 2MASS or Spitzer images of V39 are available, and Galex does not detect V39 in
either the FUV or the NUV bands. Thus the data available for this work extend from the U to the K band.

\section{Spectral morphology}
\label{spec}

\begin{figure*}
     \includegraphics[angle=0,height=6cm,width=12cm]{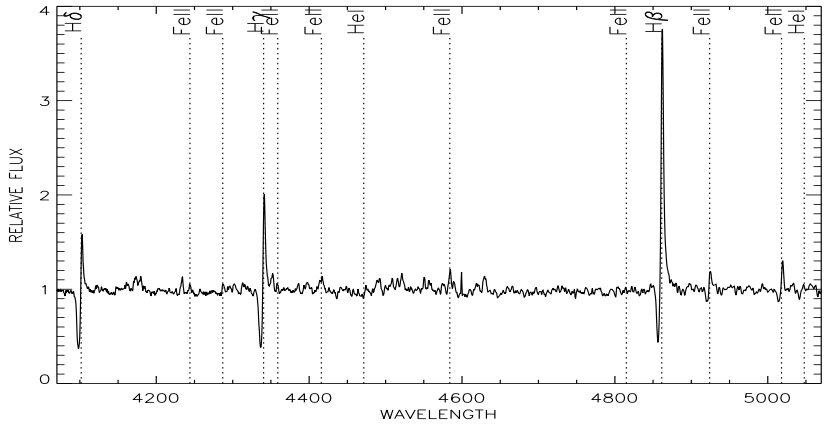}
     \caption{Blue spectrum of V39. The P-Cygni profiles of H$_\gamma$, H$_\beta$,
     and FeII] 4924, 5018 can be easily recognized. Note the absence of HeI lines. Wavelength is given in \AA.}
	\label{spectrumb}
  \end{figure*}

\begin{figure*}
     \includegraphics[angle=0,width=12cm,height=6cm]{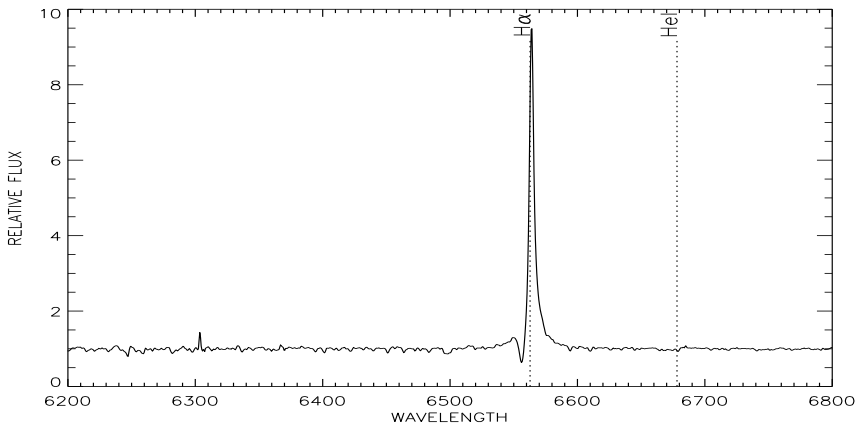}
     \caption{Red spectrum of V39. Wavelength is given in \AA.}
	\label{spectrumr}
  \end{figure*}

\begin{figure}
   \resizebox{\hsize}{!}{\includegraphics[angle=0,width=12cm,height=6cm]{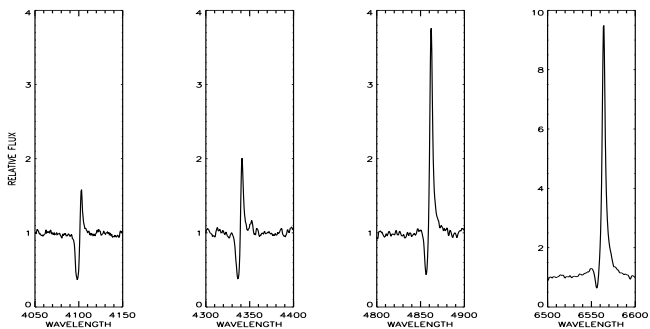}}	
     \caption{Balmer lines (H$_\delta$ to H$_\alpha$) in the spectrum of V39. Note the scale change in the y-axis for H$_\alpha$. Wavelength is given in \AA.}
	\label{balmer}
  \end{figure}

\begin{figure}
   \resizebox{\hsize}{!}{\includegraphics[angle=0]{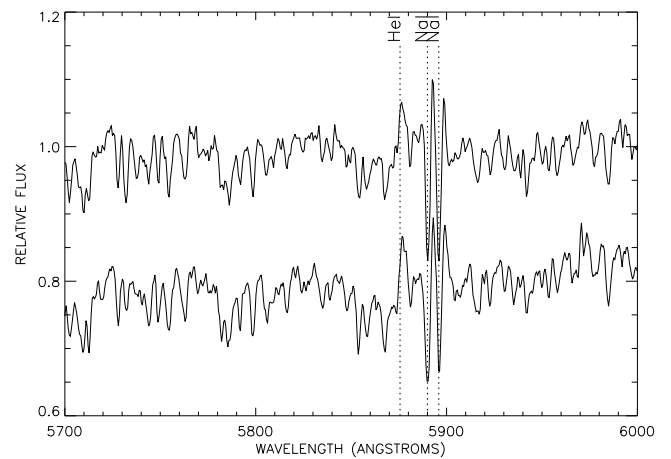}}	
     \caption{Comparison of a region of IC1613-LBVc-1 observed in the red (upper) and blue (lower) spectrograms. 
     The \ion{He}{i} and \ion{Na}{i} line positions have been marked.}
	\label{comp_br}
  \end{figure}

 Figures~\ref{spectrumb} and~\ref{spectrumr} show the observed coadded blue and red spectra of V39, respectively. The absence of 
 [\ion{O}{iii}] or \ion{He}{i} emission lines indicates that there are no contamination problems with nebular lines.  
 The spectrum is dominated by the strong
 P-Cygni profiles of the Balmer lines. They are shown in detail in Fig.~\ref{balmer}, where we can see the strong increase in the
 emission profile through the series. The second outstanding characteristic is the relatively strong P-Cygni profiles of the
 \ion{Fe}{ii} lines at $\lambda\lambda$ 4924, 5018 and 5168. The first two are very close to the \ion{He}{i} lines at $\lambda\lambda$ 4921, 5016.
 However, the fact that other \ion{He}{i} lines are much weaker or even absent, together with the presence of
 \ion{Fe}{ii} $\lambda$5168,  indicates that these are \ion{Fe}{ii} lines. The only \ion{He}{i}
 line seen in the spectrum is \ion{He}{i} $\lambda$5876. This line is present in the red and blue spectrograms, which overlap in a 
 wide wavelength region. In Fig.~\ref{comp_br} we plot part of this overlapping
 region, which shows a weak P-Cygni profile of \ion{He}{i} $\lambda$5876 in both spectra. The presence of this
 \ion{He}{i} line is very important as it can be used to constrain the stellar temperature and
 He content. 
 
 As can also be seen in Fig.~\ref{comp_br}, many weak features are present in both spectra.  This shows that
 these features are real and not the product of a noisy spectrum. The strong absorption lines redwards 
 from \ion{He}{i} are the \ion{Na}{i} D lines at $\lambda\lambda$5890, 5896. The fact that they fall
 exactly at their laboratory position after barycentric and systemic velocity corrections indicates that they belong to IC 1613, not
 to the Galaxy and that they are at rest with respect to the star. We also notice a weak redshifted
 emission component for both lines, which can initially be 
 identified either as the emission component of a P-Cygni profile or as the corresponding \ion{Na}{i} sky line at the observer's 
 rest frequency (therefore displaced by the IC1613 systemic velocity in the stellar spectrum). 
 Checking individual exposures, we realized that in two of them
 we could not completely remove the sky lines, and they appear as weak redshifted emission components in the
 stellar spectrum. 
 
\begin{figure*}
     \includegraphics[angle=90,width=15cm,height=8cm]{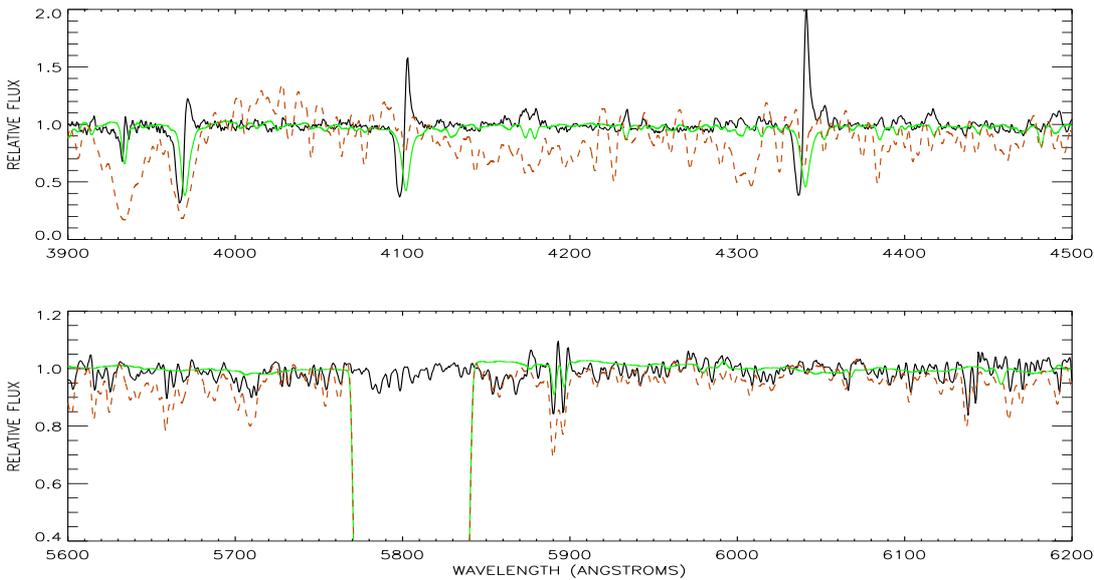}
     \caption{Comparison of the blue (above) and red (below) spectral ranges of V39 (black solid line) with the spectra of HD 62623 (A3I, green) 
     and HD 117440 (G9I, red, dashed).}
	\label{sptypes}
  \end{figure*}

 The third characteristic is that the observed spectral type seems to correspond to later types when one moves from
 blue to red spectral ranges. This is shown in Fig.~\ref{sptypes}, where we compare V39 with an A3I star and a G9I star, 
 extracted from the UVES POP
 database \citep{bagnulo03} and degraded to VIMOS resolution. Apart from the difference in line intensities (which can be
 partly attributed to the difference in metallicity) we see that V39 follows more the A3 spectrum in the blue range and 
 the G9 in the red one. Note that this is not an attempt to classify the star, but just an illustration of the spectral differences.
 
 As V39 is a photometric variable, we would expect the spectrum to also vary periodically. However, changes in our 
 individual spectra are
 small and mostly compatible with difficulties in the sky line reduction or within noise limits.  For example, 
 the variability we can see in the \ion{Fe}{ii} profiles is modest. Moreover, no significant radial velocity variability 
 is detected (although the relatively low resolution and modest S/N of the individual spectra limit our
 ability to detect radial velocity changes). As an example we can examine the line profiles of H$_\alpha$,
as this line is known to be very sensitive to conditions in the atmospheric layers. In Fig.~\ref{Hatime}
we plot the H$_\alpha$ profiles observed during seven days 
(i.e. covering a quarter of the 28 days period or half the 14 days period), and we see that
only small changes are present.
H$_\beta$, observed during a time span of 30 days, displays similar small changes.
Figure~\ref{Haew} displays the equivalent widths measured in the individual H$_\alpha$
spectra for reference. We see again that the variations are small. Changes in H$_\beta$ are 
once more similar, although because of the smaller values and lower SNR, the relative changes 
appear a bit larger. 
 
\section{The nature of V39}
\label{nature}

The spectrum described in the previous section resembles partly that of an LBV candidate. The strong emission and moderate maximum velocity
of the P-Cygni profiles of the Balmer lines point to an intense, dense and slow wind, like that of a supergiant. 
At the same time, the \ion{Fe}{ii} lines and the lack of a strong He spectrum points to an intermediate temperature.
The cool features in the red spectrum indicate some degree of contamination by a secondary source
that however shares the radial velocity of the blue spectrum. In addition, V39 displays periodic or quasi-periodic photometric variations that are
not expected in an LBV. For that reason we analyze here different alternative scenarios for V39
using the previously known facts about V39 given in the introduction and the spectral description from previous section. 

\begin{itemize}

\item[i)] {\it a Galactic W Vir star plus an IC 1613 red supergiant.} Following Antonello's et al. ideas, 
\citet{mantegazza02} obtained two periods of
14.341 and 1118 days. They offered a solution for V39: the superpositon along the line of sight of a Galactic W Vir star,
a Population II Cepheid pulsating with a period of 14.341 days, and a red supergiant in IC 1613 that varies with
a period of 1118 days. The alignment has to be in a way that both stars contribute to the light we receive at Earth.
This would agree with a composite spectrum and,
in spite of the unprobable configuration, this interpretation has an additional interest: it implies that the W Vir star should 
be at a distance of 115 Kpc. This would mean that it is the most distant Galactic star known and would raise the question 
about the true extension of the Galactic halo.

However, our higher resolution spectrum is not consistent with this classification. Firstly, the barycentric plus systemic velocity 
correction is consistent with the rest of the stars observed in IC1613, both in the blue and red parts of the spectrum,
and therefore consistent with V39 being a member of IC 1613. Moreover, there is no radial velocity shift between
the red and blue spectra. And secondly, the timeseries of the individual spectra 
before coadding them shows that the P-Cygni profiles of the Balmer lines vary only slightly (see Fig.~\ref{Hatime}), 
whereas Balmer lines in W Vir stars vary strongly along their cycles and never show P Cygni profiles (see Fig. 3 of \cite{lebre92}).
 

\begin{figure}
   \includegraphics[angle=0,height=8cm,width=8cm]{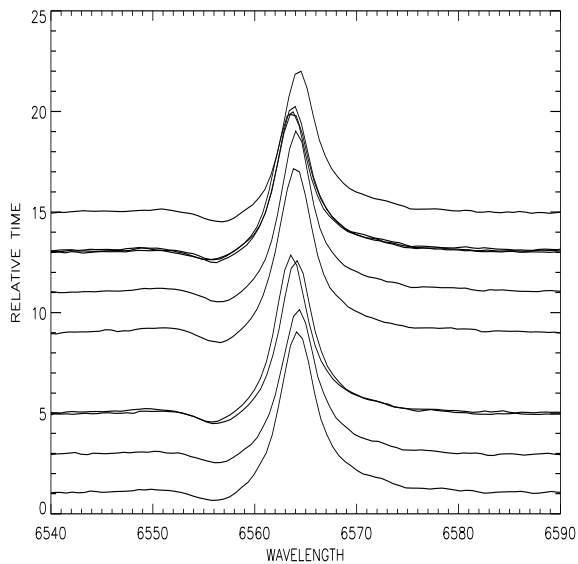}
   \caption{Time series of H$_\alpha$ spectra of V39. Relative fluxes have been displaced
     according to the time difference with respect to the first observed spectrum. Units in the ordinate axis are half-days.
     The time difference between the first and last spectra is seven days. Wavelength is given in \AA.}
	\label{Hatime}
  \end{figure}


\item[ii)] {\it A close binary system.} Another possibility to explain the periodic variations is V39 being actually a binary system, composed for example
of a supergiant dominating the spectrum and a hotter companion. We could assume the binary system motion being
responsible for the short period, whereas the supergiant would be responsible for the long-period variability. 
However, this explanation is not supported by the data.
The short period of about 14 or 28 days is too short to be caused by binary motion. It would imply
separations smaller than 75 or 150 R$_\odot$ respectively (which are not compatible with the presence of a
red supergiant) and large radial velocity variations that are not detected. 

\item[iii)] {\it A wide binary system.} Alternatively, we could attribute the
long period with its small magnitude changes to the binary motion, and the short period with its large 
magnitude changes to the primary (brightest) star.
The long period of 1120 days would imply a larger separation than the short one, about 1800 R$_\odot$, and modest
radial velocity variations (of 30-50 km s$^{-1}$, which means a maximum separation of about 60-100 km s$^{-1}$
every 1.5 years). At the distance of IC 1613, this spatial separation is too small to be detected, 
and our spectral resolution is too low to detect the implied velocity changes in only 40 days (that
is the time span covered by our observations). Assuming that the light curve of the long period is 
due to binary motion, we would attribute a change of $\Delta$B= 0.3 to contamination
by the secondary (from Figure 11 of \citet{antonello99}), corresponding to a star with m$_B \geq$ 20
or M$_B \geq$ -4.2, which is too low for any supergiant, implying an
object with a small radius and a blue color. Depending on the exact relative magnitudes, this may or may not be
detectable in the composite spectrum. However, the major photometric changes 
should come from the supergiant and are thus inconsistent with no spectroscopic changes during the
photometric period.

\item[iv)] {\it Contamination by a nearby object.} This situation is similar to that of a binary system, but without the
restrictions on object separation. Therefore, we will only consider the case that the short period
variations are due to a nearby star or binary system, close enough along the line of sight to the star 
that dominates the observed spectrum
(i.e., much in the line of Mantegazza et al., but now with both objects belonging to IC 1613).
In this case the spectrum would be dominated by a red supergiant responsible for the long period and the
second object would be a star or stellar system in IC1613 responsible for the short period variations,
either because it is an eclipsing binary system or because it is a pulsating star. 
However, as the magnitude changes are considerable, the nearby object should be relatively bright,
which restricts the possibilities. If, for example,
we attribute the $\Delta$B= 1.0 mag. and the $\Delta$V= 0.5 variations to this nearby object
(using Figs. 11 and 3 of \citet{antonello99} and \citet{mantegazza02}, who have already substracted
the long-period effect) the observed photometric changes are compatible with the supergiant dominating the
blue spectrum only during less than half of the period. As we have observed our object for more than one
period (when we take blue and red spectra into account) without significant spectral changes, we discard this scenario.

\item[v)]{\it A star surrounded by a hot thick disk.} In this scenario the star is surrounded by a thick disk that 
emits at a characteristic temperature of 5000 K, producing the spectral features in the red part
of the observed spectrum. As we do not expect a star of about 10000 K to show relevant spectral features in
the 6000 \AA~ region at our resolution (see Fig.~\ref{sptypes}), 
we can attribute them to the disk even if it does not completely dominate the system light at these
wavelengths (of course, this is not true for some particularly intense lines in the spectrum of the star, like
H$_\alpha$; conversely, the stellar Balmer lines in the blue could also be affected by the disk lines). The 
blue spectrum would be produced by the star itself and represent a hotter temperature. 
The photometric changes would be due to disk precession (at least partly) changing the aspect angle of the system.
The relative contributions of star and disk would vary periodically,
but always producing the same combined spectrum, except for small spectral line variations (in absence of
precession, photometric changes without spectroscopic ones would imply a change in the stellar
bolometric luminosity, which also argues against a single star explanation). 
In this scenario, photometric changes are wavelength dependent, as observed, as the disk becomes
thinner at larger wavelengths. Details of such a scenario have still to be worked out to show that
the large photometric changes are actually compatible with the small spectroscopic changes, but
it is consistent with present data. Moreover, the possible presence of a disk is reinforced by the large 
reddening of V39, much larger than the average one towards IC 1613, than the typical internal reddening in this galaxy,
and than that of the surroundings of V39. 
For example, \cite{lee93} give E(B-V)=0.02 towards IC1613, while \cite{garcia09} give E(B-V)= 0.10-0.15 as the mode 
of the distribution of extinction for OB associations. Moreover,
the spectra of the three stars closest to V39 in our VIMOS mask do not
show appreciable \ion{Na}{i} lines. Therefore the reddening must be very local. However,
we have no further independent evidence supporting the disk scenario. New spectra with higher spectral resolution
to investigate for example the presence of a double peak produced by a disc would be very helpful to confirm or discard it. They should
be ideally obtained simultaneously with new photometric data, as the available photometry has been obtained 
in the course of decades, by different authors using different filters.

\end{itemize}

\begin{figure}
   \includegraphics[angle=0,height=6cm,width=8cm]{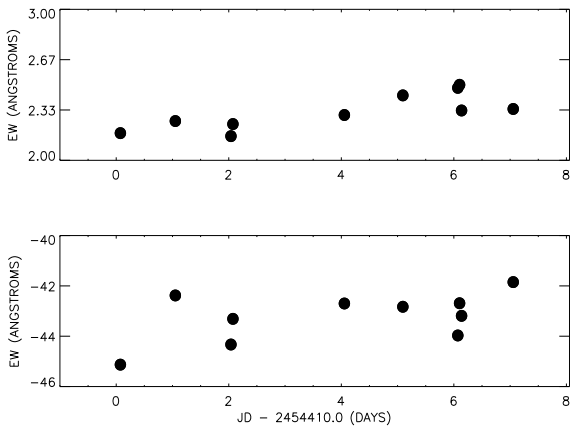}
\caption{The equivalent widths of H$_\alpha$ (in \AA) of the individual spectra available. Upper plot:
absorption profile; lower plot: emission profile. Changes do not show any clear
periodicity, nor are they larger than a few percent.}
\label{Haew}
  \end{figure}









\section{Spectroscopic analysis}
\label{param}


We have concluded that the observed spectrum of V39 is best interpreted as that of a BA-type supergiant
surrounded by a hot, thick disk precessing around it. The star itself could be an LBV candidate,
attending to the strong P-Cygni profiles shown in the Balmer and \ion{Fe}{ii} lines.
To be classified as an LBV the star should show outbursts and photometric and spectroscopic variations. There are no
known outbursts of V39, but part of the periodic or quasi periodic photometric variations could be intrinsic
to the star. This possibility has been explored by \cite{onifer08}, who show that it is possible for LBVs 
at low metallicity and large He abundances to develop quasi-stable pulsational instabilities with periods of about
one month due to a $\kappa$-effect mechanism caused near the iron bump opacity. 

To further investigate the possibility that the star is an LBVc we have carried out a quantitative analysis of 
the spectrum to place it in the theoretical Hertzsprung-Russell Diagram and to compare its position with
those of other stars. For this purpose, we performed a model atmosphere analysis of the spectrum of V39 obtained after 
co-adding the individual spectra, as described in Sect.~\ref{obs} (19 blue spectra and 10 red spectra). 

The stellar parameters were determined by means of the CMFGEN model atmosphere code \citep{hillier98}. Details of the
version and atomic models used  are given in \cite{najarro09}.
In Tables~\ref{modpar} and \ref{modpar2} we list the main model parameters and the synthetic photometry, obtained by
integrating the emergent spectral distribution weighted by appropriate filter values.

As no ionization equilibrium could be used for T$_{\rm eff}$ determination, 
we have adopted a He abundance characteristic of LBV stars, namely H/He= 2.5 (\cite{najarro97},
\cite{najarro09}). 
T$_{\rm eff}$ was then determined paying attention to the presence of \ion{Fe}{ii} P-Cygni lines, the probable presence of \ion{He}{i} 
$\lambda$5876 and the absence of other \ion{He}{i} lines. 
We choose to fit the H$_\alpha$ line as this is the strongest spectral feature, and should
be dominated by the mass-losing star in spite of the contamination of the red spectrum.

The radius (and therefore the luminosity) was derived from the K-band magnitude, as this is the wavelength with the
lower reddening and for which the magnitude varies less (we note that the changes in brightness decrease with increasing wavelength). 
We assume here that in a star+disk scenario the stellar contribution dominates at this wavelength.
To obtain a reddening estimation we used the photometry given by \cite{udalski01}, as it covers a time span similar
to that of our spectroscopic observations (40 nights in our case, 51 in that of Udalski et al.; note that using our 
own photometry given in \cite{garcia09} is less appropriate, as it was obtained in one single night). 

\begin{figure}
   \resizebox{\hsize}{!}{\includegraphics[angle=0]{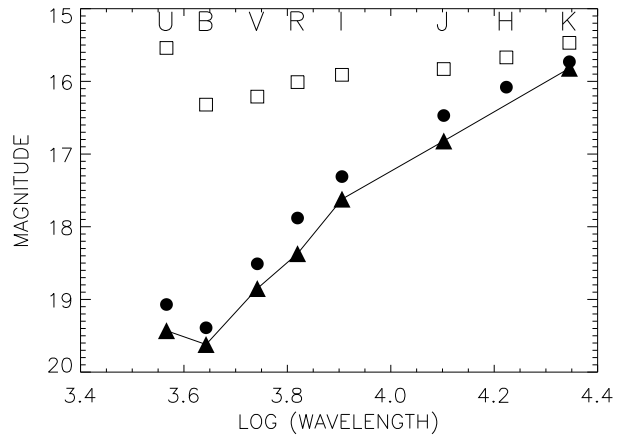}}	
     \caption{Comparison of observed magnitudes of V39 listed in Table~\ref{stardata} (triangles) with those of our model, both
     intrinsic (open squares) and reddened (circles). Wavelength is given in \AA~ and the photometric bands
     are indicated on top of the plot. See text for details.}
	\label{magnitudes}
  \end{figure}

We proceeded in the following way: from the unreddened (V-I) color obtained from the model and the observed (V-I) from
\citet{udalski01} we derived (A$_{\rm v} - A_{\rm I}$)= 0.92. Using then (A$_{\rm I}$/A$_{\rm v}$)= 0.601 and 
(A$_{\rm k}$/A$_{\rm v}$)= 0.112, we finally got A$_{\rm k}$= 0.26 and K= K$_{\rm model}$ + 0.26= 15.73 (compared
with the observed K= 15.82). With these values we obtained I= 17.30 and V= 18.53, which is slightly brighter than
the values observed by \citet{udalski01}, namely I= 17.62 and V= 18.85. 
This is more clearly seen in Fig.~\ref{magnitudes}, where we plotted the observed magnitudes 
from Table~\ref{stardata} and the predicted magnitudes from our spectroscopic analysis, both intrinsic and reddened
(for the observed magnitudes we used those from \cite{udalski01} for the reasons cited above,
but we used our own value for the B band, as the values from \cite{sandage71} are old measurements that had to
be calibrated through a complicated process (see Sect. III.c in Sandage's paper)). We see that the model reproduces
the observed data well after including extinction, although an extra absorption would be needed in all bands relative
to the K-band. This can be due to a number of factors: a slightly different reddening law at low Z, extra emission
in the K band, or extra absorption at shorter wavelengths, produced by a disk, or a combination of all of them.

\begin{figure}
   \resizebox{\hsize}{!}{\includegraphics[angle=0]{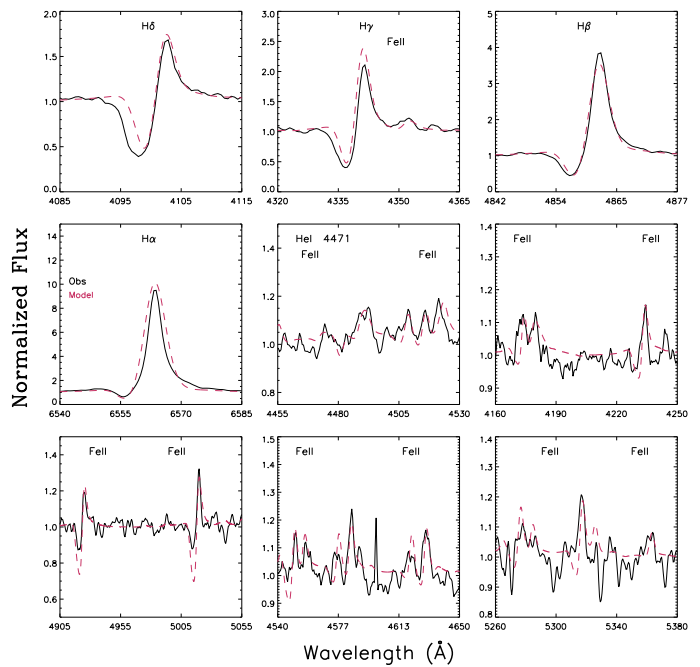}}	
     \caption{Fit to the observed spectrum of V39 calculated with CMFGEN. Stellar parameters are given in Table~\ref{modpar}.
     }
	\label{spcomp}
  \end{figure}

In Fig.~\ref{spcomp} we show the comparison of selected lines in the observed spectrum with the CMFGEN model, 
after being convolved with the appropriate instrumental resolution. Although abundances have not
been determined, implying that the fit cannot be perfect, we see that the general behavior is well reproduced by the model.
No effort to determine the rotational velocity has been carried out, as the resolution is modest. Parameters in Table~\ref{modpar} have
been adopted as the final stellar parameters.

We see that the fit to the observed spectrum is quite good, although 
there is a trend for the absorption part of the higher Balmer lines to be broader than the model calculation. This cannot be 
attributed to terminal velocity uncertainties as the lower members fit well (note that Galactic stars would have spectra redshifted with
respect to IC 1613 stars). In a star+disk scenario, this could be due to the differential absorption produced by
the disk in the series members combined with the different levels of disk continuum. We also emphasize that 
the absorption part of some \ion{Fe}{ii} P-Cygni profiles (\ion{Fe}{ii} $\lambda\lambda$4924, 5018) is predicted too strong
as compared to observations. This is not solved by changing the stellar parameters, and thus we rely on the emission peaks.
These points might indicate that the density structure of V39 is not as simple
as described by our spherically symmetric, one-dimensional models. The presence of a disk is not the only
possibility: wind clumping, for example, was not considered.  
As is well known (see \citet{puls08}), the presence of clumping may modify the mass-loss rates 
derived from the H$_\alpha$ line, for which we obtain a value that is higher than expected for a star
of the given luminosity and metallicity. Whether this represents a challenge for the theory of radiatively driven
winds cannot be said at the present stage. 

We derive a rather typical terminal velocity for an LBVc, 390 km s$^{-1}$, and thus find no compelling evidence that 
v$_\infty$ increases with decreasing metallicity, as suggested by \citet{izotov09}. Increasing the terminal
velocity results in too broad absorption in H$_\alpha$ and H$_\beta$ and does not help further. We note that \citet{drissen01} also
found a modest terminal wind velocity for the low-metallicity LBV NGC 2363-V1 (290-325 km s$^{-1}$). 

We have not determined metal abundances, but we have constrained the Fe abundance of V39, 
which is an important parameter, as Fe is the main driver of the stellar wind.
We obtained (Fe/Fe$_\odot$)= 0.2, which agrees with the value by 
\citet{taut07}, who obtained a mean iron abundance of [Fe/H]= -0.67$\pm$0.09 from the analysis of three M-supergiants in IC 1613, 
which corresponds to 0.21 Fe$_\odot$ and the present value of Z= 0.004 ([Fe/H]= -0.7) derived by \cite{skillman03} from isochrone 
fitting to the young stellar population of IC 1613.
All three metallicities are higher than the metallicity derived from nebular and stellar O, for which we find in the literature 
values between 7.60 and 7.90 (see \citet{lee03}, \citet{bresolin07}, among other results), which corresponds to 
0.08 - 0.16 times the solar oxygen abundance (assuming the solar O abundance from \cite{asplund09}). The fact that the 
B supergiants and \ion{H}{ii} regions give consistent O abundances and that the same consistency is found for the Fe
abundances from (one) hot , (three) cool supergiants and the young stellar population is a remarkable result, 
which supports the conclusion that the [$\alpha$/Fe] ratio in IC 1613 is slightly lower than solar.

Given the number of parameters and the complexity of the models, a proper 
  error study is not possible. Within the parameter domain the star is
   placed and for the assumed H/He ratio, the errors in
   the stellar temperature are narrowed down to $\pm$1000K given the presence
   of weak \ion{He}{i} 5876 and absence of other \ion{He}{i} lines. The errors in radius and
   luminosity are dominated by the photometry and reddening uncertainties, while
   the error in radius dominates the error in mass-loss (transformed radius).
   Because of  the large number of \ion{Fe}{ii} lines in the spectra of V39 the
    \ion{Fe}{} abundance can be well constrained once the T$_{\rm eff}$ and $\dot{\rm M}$ have been
    determined. Thus, 0.15 dex can be regarded as a conservative error on the \ion{Fe}{} abundance.

\begin{table*}[htdp]
\caption{Stellar parameters obtained in this work for V39 with CMFGEN and for similar stars taken from the literature.}
\centering
\begin{tabular}{l c c c c c c c c l}
\hline \hline
Star & T$_{rm eff}$ & log (L/L$_\odot$) & R/R$_\odot$ & M$_{\ast}$ & He/H & $\dot{\rm M}$ & V$_\infty$ &  (Fe)/(Fe)$_\odot$  & Reference\\
        &(K)           &                             &                            & (M$_\odot$)     &          &(M$_\odot$ yr$^{-1}$) & (km s$^{-1}$)     &      &                \\
\hline
V39             & 9260 &  5.34& 183 & 25 & 0.4 & 3.0$\times$10$^{-5}$ & 390 &   0.2 & this work \\

HD 160529 & 8000 & 5.60 & 330 & 13 &       & 1-3$\times$10$^{-5}$ &  120  &       & \citet{sterken91} \\
R 110         & 7600 & 5.46 & 310 & 10 &       & 1$\times$10$^{-6}$ &      80  &         & \citet{stahl90}\\
R 40           & 8700 & 5.61 & 280 & 16 &       & 8$\times$10$^{-6}$ & 125 &             & \citet{szeifert93}\\
\hline
\end{tabular}
\label{modpar}
\end{table*}
\begin{table*}[htdp]
\caption{Synthetic photometry for the model given in Table~\ref{modpar}. Values are unreddened.}
\centering
\begin{tabular}{c c c c c c c c c}
\hline \hline
M$_{\rm v}$ & U & B & V & R & I & J & H & K \\
\hline
-8.06            & 15.54 & 16.32 & 16.21 & 16.01 & 15.91 & 15.83 & 15.67 & 15.47\\
\hline
\end{tabular}
\label{modpar2}
\end{table*}
  

\section{Discussion}
\label{discus}
\begin{figure}
   \resizebox{\hsize}{!}{\includegraphics[angle=0]{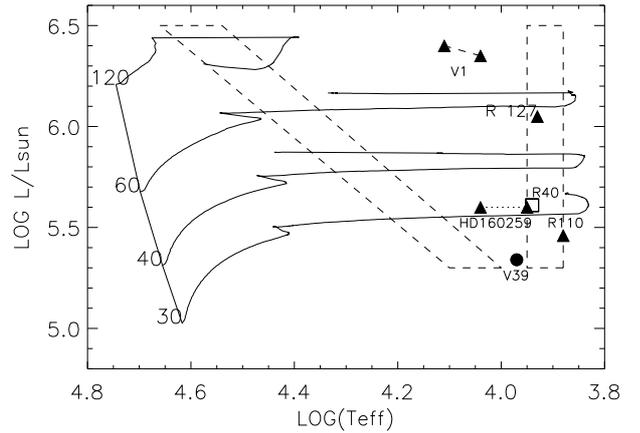}}	
     \caption{The position of V39 in the HR Diagram (filled circle). For comparison we have also plotted (filled triangles) the positions of 
      NGC 2363-V1 and R127 during outburst, as well as the low luminosity LBVs HD 160529 (during two stages connected by a dotted line), 
      and R110. Also plotted is the position of the SMC LBV R40 (with an open square, to avoid confusion). 
      We also marked the region occupied by LBVs during outburst (vertical box) 
     and the one occupied by LBVs during visual minimum (slanted strip, both from \cite{smith04}). The tracks correspond to evolutionary models for the SMC 
     including rotation (with initial velocities of 300 km s$^{-1}$, \cite{meynet05}). Initial track masses are given by the 
     numbers left from the ZAMS.}
	\label{hrd}
  \end{figure}

In Fig.~\ref{hrd}
we have plotted evolutionary tracks including rotation for the SMC metallicity \citep{meynet05}
together with the position of V39 from the model parameters and some known LBVs. 
From its position in the diagram we can estimate the 
initial stellar mass to be $\sim$25 M$_\odot$. According to the evolutionary tracks,
the star should be in a redward excursion, without possible return to the blue. Because of the low
metallicity and consequently weak wind, the mass-loss in the model up to its present stage 
is very small, and therefore the present mass predicted by the models is not very different from the initial mass. 

However, the star may have suffered previous episodes of mass loss, and the possible presence of 
a disk may be an indication of it. At the derived stellar temperature, the \ion{Na}{i} lines are expected to be 
much weaker than observed. The spectral type should be at least F5, according to the stellar atlas of \citet{dansk94} 
(that contains Galactic stars) to have comparable stellar lines. Therefore we expect the \ion{Na}{i} D lines to be formed
in the circumstellar material, and we conclude that their presence is consistent with (but not a proof of!) previous episodes of mass loss. 



The derived luminosity is at the lower limit of the known LBV and LBV candidates. 
The absolute visual magnitude we derive is  M$_{\rm v}$= -8.06, which is modest
for an LBVc, but again within the expected range. The position of V39, close to the low luminosity end of known LBVs, 
adds interest to this object, as the number of
known low-luminosity LBVs (or LBV candidates) is very small. In Fig. 4 of \citet{smith04} we see only seven objects in this region
of the HRD, including
both LBVs and LBV candidates, belonging to the Milky Way, the LMC and the SMC. 
In Fig.~\ref{hrd} we have also marked the position occupied by 
NGC 2363-V1 \citep{drissen01} and R127 during outburst, and the low-luminosity LBVs HD 160529 (Galactic, \citet{sterken91}),
R110 (or HD 269662 in the LMC, \citet{stahl90}) and R40 (or HD 6884 in the SMC, \citet{szeifert93}). 
While the first two are much more luminous than V39, the latter ones are closer to it, and we are tempted to 
conclude that they are in a similar stage, although a comparison of the spectrum of V39 with those
available in the literature for HD 160529 \citep{chentsov03}, R110 \citep{stahl90} or R40 \citep{szeifert93} results in clear differences:
the P-Cygni profiles of Balmer and Fe lines are
more pronounced in V39, which could indicate that the star is closer to its Eddington limit. This would be consistent with
V39 having ejected a large amount of matter, indicated by its intense reddening (the reddening towards HD 160295 is more intense,
but this is a Galactic star at low Galactic latitude, {\it l}= -1.73) and having the largest mass-loss rate, in spite of its lower
metallicity. Whether this is related to the main difference, namely the strong, short-period photometric variations
present in V39, cannot be clarified with the present data. However, it is clear that the derived mass-loss rate is very
high for a star of the given luminosity and metallicity.
Table~\ref{modpar} gives the main parameters of the stars. Note that for HD 160529, R110 and R40 values
were not obtained through a detailed spectral fit but with a combination of plane-parallel and wind models (details can
be found in the cited references).

Alternatively, V39 could represent a high-luminosity version of supergiant B[e] stars (sgB[e], see \cite{lamers98}), particularly if
we consider that our luminosity might be slightly overestimated (we have considered all K-band emission as stellar). An argument 
supporting this possibility is the indication by \cite{mirosh07} that these stars are more luminous at lower metallicities (but
given the uncertainties in the luminosity determination and the low sgB[e] statistics we have to be cautious with this particular argument).
When attributing the cooler spectral features to the disk and the main photometric variations to disk precession, observed features
would be compatible both with a LBVc and a sgB[e] nature. The main difference would be the lower luminosity expected for
sgB[e] stars and its larger infrared excess at long wavelengths (because of dust in the circumstellar material). There are no 
Spitzer or IRAS data providing evidence of large IR excesses for V39, 
and therefore we prefer its classification as an LBV candidate, but without
completely discarding a possible sgB[e] nature which, on the other hand, may simply be different evolutionary stages of the same star.
However, we need some force causing disk precession. This can be due either to a non-stable disk or to perturbation from 
some companion. In the first case, the disk should have formed recently (otherwise friction would stop precession) or even
being still in formation (the high derived mass-loss rate supports this possibility). The second case is not discarded by our
discussion in Sect.~\ref{nature}, as the companion should only be responsible for the small perturbation causing precession, and not
for the photometric changes.

%


\section{Conclusions}
\label{conc}

We have presented intermediate resolution (R=2050-2150)
spectroscopic observations of the variable star V39 in IC\,1613 secured with VIMOS at the VLT.
The final spectrum is the combination of 19 blue spectra and 10 red spectra and is characterized by
i) strong P-Cygni profiles in the Balmer lines and \ion{Fe}{ii} lines; ii) an hybrid character, more consistent
with an early type supergiant (B-A) in the blue and a late type (G) in the red. All spectral features are
consistent in radial velocity with the star belonging to IC 1613. 

The individual spectra do not show significant radial velocity variations in spite of the time span covered, which
is longer than one photometric period (during which photometric changes are larger than one magnitude
in B and decrease with wavelength), and show only moderate spectral line
variations. Therefore we have discarded scenarios including a binary system, specially those demanding
large radial velocity variations. We also discarded the scenario
proposed by \cite{mantegazza02} (a Galactic W Virginis star plus an IC 1613 red supergiant)
because it is not consistent with the system belonging to IC1613 and the observations do not show
the spectral characteristics and changes of a W Virginis star. We propose a scenario consisting of an
early (B-A) type star surrounded by a hot ($\sim$ 5000 K) thick disk precessing around it. Although
this scenario can potentially explain the photometric changes and its wavelength dependence
as well as the lack of significant spectroscopic changes, details have to be worked out (for example,
the cause of disk precession) and independent
evidence (like for example the presence of double peaks in spectral lines) has still to be obtained.

The analysis of the spectrum under the assumption that this is a single star indicates a low-metallicity (Fe $\sim$ 0.2 Fe$_\odot$,  
  from the iron P-Cygni lines),   hot evolved star with a strong wind. The agreement of this result with other spectroscopic Fe or 
isochrone fitting global metallicities (\citet{taut07}, \citet{skillman03}) on the one hand and the stellar and nebular O abundances
on the other (\citet{bresolin07}, \citet{lee03}) strongly points to a low [$\alpha$/Fe] ratio in IC 1613 as compared to solar values, 
although more work will be needed to confirm this possibility.  

From its position in the HRD the star lies
in the low-luminosity edge of LBV and LBV candidates and its progenitor may have had a mass of about 25 M$_\odot$
(which would be close to the present mass). The derived mass-loss rate is higher than expected for a star of the
given luminosity and metallicity, and it may represent a challenge to theory.
The terminal velocity is consistent with other LBVc, and we could not confirm the suggestion by \citet{izotov09} 
that it increases at low metallicities. Our result agrees with that of \citet{drissen01} for NGC 2363-V1.
If confirmed, V39 would be the resolved LBV candidate located in the most metal poor environment known to date. 
Alternatively, the star could be a high-luminosity version of a sgB[e]. This second alternative would imply some
significant dust formation around V39 (which we could neither confirm nor reject) and is not inconsistent with the star being an LBV at some stage in its
evolution.

The combination of the low abundances
with the high mass-loss rate raises the issue of how the winds of evolved stars behave at low metallicities (like those of LBV and
sgB[e] stars) and whether they are driven by line radiation, pulsations or continuum radiation playing a stronger role than in Galactic stars. 
Our data are not sufficient to explore this question in detail. Higher resolution spectra, allowing us to dissentangle the
contributions of the star from that of possible circumstellar material, disk or faint companion are needed. 

Whether V39 is an LBVc or a new transient phase of massive star evolution, an LBV impostor, is an important question. If
the former, then it will hold the record as the lowest metallicity LBVc known which has important implications for stellar evolution
(as described in the introduction). If the latter, then understanding what this phase might represent has great intrinsic interest,
in addition to raising the issue of whether or not there may be other LBV impostors which are undiscovered due to a
lack of adequate data, like a light curve. Unraveling its nature is therefore of great importance.


\begin{acknowledgements}

The authors thank I. Negueruela and S. Simon-Diaz for valuable and inspiring discussions.
We also thank the referee, Dr. Laurent Drissen, for many positive and valuable comments
during the revision of the manuscript.
This project has been supported by Spanish grants AYA2007-67456-C02, AYA2008-06166-C03-01 and
AYA2008-06166-C03-02 and was partially funded by the Spanish MICINN under the Consolider-Ingenio 
2010 Program grant CSD2006-00070: First Science with the GTC 
(http://www.iac.es/consolider-ingenio-gtc).
We acknowledge the use of UVES POP data obtained under ESO DDT Program ID 266.D-5655

\end{acknowledgements}


\bibliographystyle{aa}

\begin{thebibliography}{}
\bibitem[Antonello et al. (1999)]{antonello99} Antonello, E., Mantegazza, L.,  Fugazza, D., Bossi, M. \& Covino, S., 1999, A\&A, 349, 55


\bibitem[Asplund et al. (2009)]{asplund09} Asplund, M., Grevesse, N., Sauval, Scott, P., 2009, ARA\&A, 47, 481

\bibitem[Bagnulo et al. (2003)]{bagnulo03} Bagnulo, S., Jehin, E., Ledoux, C. et al., 2003, Messenger 114, 10

\bibitem[Bresolin et al. (2007)]{bresolin07} Bresolin, F., Urbaneja, M.A., Gieren, W., Pietrzy\'nski, G. \& Kudritzki, R.P., 2007, \apj, 671, 2028

\bibitem[Chentsov et al. (2003)]{chentsov03} Chentsov, E.L., Ermakov, S.V., Klochkova, V.G. et al.,  2003, A\&A, 397, 1035

\bibitem[Clark et al. (2005)]{clark05} Clark, J.S., Larionov, V.M. \& Arkharov, A., 2005, A\&A, 435, 239

\bibitem[Dansk \& Dennefeld (1994)]{dansk94} Dansk, A.C. \& Dennefeld, M., 1994, \pasp, 106, 382

\bibitem[Dolphin et al. (2001)]{dolphin01} Dolphin, A.E., Saha, A., Skillman, E.D. et al., 2001, \apj, 550, 554

\bibitem[Drissen et al. (2001)]{drissen01} Drissen, L., Crowther, P.A., Smith, L.J. et al., 2001, \apj, 546, 484

\bibitem[Lebre \& Gillet (1992)]{lebre92} L\`ebre, A. \& Gillet, D., 1992, A\&A, 255, 221 

\bibitem[Garcia et al. (2009)]{garcia09} Garcia, M., Herrero, A., Vicente, B. et al., 2009a, A\&A, 502, 1015


\bibitem[Gonz\'alez-Delgado et al. (1994)]{gonzalez94} Gonz\'alez-Delgado, R.M., P\'erez, E., Tenorio-Tagle, G. et al., 1994, \apj 437, 239

\bibitem[Hillier \& Miller (1998)]{hillier98} Hillier, D.J. \& Miller, D.L., 1998, \apj, 496, 407


\bibitem[Izotov et al. (2007)]{izotov07} Izotov, Y.I., Thuan, T.X., \& Guseva, N.G., 2007, \apj, 671, 1297

\bibitem[Izotov \& Thuan (2009)]{izotov09} Izotov, Y.I. \& Thuan, T.X., 2009, \apj, 690, 1797

\bibitem[Karachentsev et al. (2002)]{kara02} Karachentsev, I.D., Dolphin, A.E., Geisler, D. et al., 2002, A\&A, 383, 125

\bibitem[Kotak \& Vink (2006)]{kotak06} Kotak, R. \& Vink, J.S., 2006, A\&A, 460L, 5

\bibitem[Lamers et al. (1998)]{lamers98} Lamers, H. J.G.L.M., Zickgraf, F.J.,  de Winter, D.,  Houziaux, L. \&  Zorec, J., 1998, A\&A 340, 117

\bibitem[Lee et al. (1993)]{lee93} Lee, M.G., Freedman, W.L. \& Madore, B.F., 1993, \apj, 417, 553

\bibitem[Lee et al. (2003)]{lee03} Lee, H., Grebel, E.K. \& Hodge, P.W., 2003, A\&A, 401, 141

\bibitem[Lu et al. (1993)]{lu93} Lu, N.Y., Hoffman, G.L., Groff, T., Roos, T. \& Lamphier, C., 1993, \apjs, 88, 383

\bibitem[Luridiana et al. (1999)]{luridiana99} Luridiana, V., Peimbert, M. , \&  Leitherer, C., 1999, \apj, 527, 110

\bibitem[Makarova \& Karachentsev (1998)]{makarova98} Makarova, L.N. \& Karachentsev, I.D., 1998, A\&AS, 133, 181

\bibitem[Mantegazza et al. (2002)]{mantegazza02} Mantegazza, L., Antonello, E., Fugazza, D., Covino, S. \& Israel, G.L., 2002, A\&A, 388, 861


\bibitem[Meynet \& Maeder (2005)]{meynet05} Meynet, G. \& Maeder, A., 2005, A\&A, 429, 581

\bibitem[Miroshnichenko (2007)]{mirosh07} Miroshnichenko, A.S., 2007, ApJ 667, 497

\bibitem[Najarro et al.(1997)]{najarro97} Najarro, F., Hillier, D.~J., \& Stahl, O.\ 1997, \aap, 326, 1117 

\bibitem[Najarro et al. (2009)]{najarro09} Najarro, F., Figer, D.F., Hillier, D.J., Geballe, T.R. \& Kudritzki, R.P., 2009, \apj, 691, 1816

\bibitem[Onifer et al. (2008)]{onifer08} Onifer, A.J. \& Guzik, J.A., in IAUS 250, eds. J. Puls, P. Crowther and F. Bresolin, p. 83

\bibitem[Owocki et al. (2004)]{owocki04} Owocki, S.P., Gayley, K.G. \& Shaviv, N.J., 2004, \apj, 616, 525

\bibitem[Petit et al. (2006)]{petit06} Petit, V., Drissen, L. \& Crowther, P.A., 2006, \aj, 132, 1756

\bibitem[Pietrzy\'nski et al. (2006)]{pietr06} Pietrzy\'nski, G., Gieren, W., Sozy\'nski, I. et al., 2006, \apj, 642, 216

\bibitem[Puls et al. (2008)]{puls08} Puls, J., Vink, J.S., \& Najarro, F., 2008, A\&ARv, 16, 209


\bibitem[Pustilnik et al. (2008)]{pustilnik08} Pustilnik, S.A., Tepliakova, A.L., Kniazev, A.Y. \& Burenkov, A.N., 2008, MNRAS, 388, L24

\bibitem[Sandage (1971)]{sandage71} Sandage, A., 1971, \apj, 166, 475

\bibitem[Skillman et al. (2003)]{skillman03} Skillman, E.D., Tolstoy, E., Cole, A.A. et al., 2003, \apj, 596, 253

\bibitem[Smith et al. (2004)]{smith04} Smith, N., Vink, J.S. \& de Koter, A., 2004, \apj, 615, 475

\bibitem[Smith \& Owocki (2006)]{smith06} Smith, N. \& Owocki, S. P., 2006, \apj, 645, L45

\bibitem[Stahl et al. (1990)]{stahl90} Stahl, O., Wolf, B., Klare, G., J\"uttner, A., Cassatella, A., 1990, A\&A, 228, 379

\bibitem[Sterken et al. (1991)]{sterken91} Sterken, C., Gosset, E., J\"uttner, A., et al., A\&A, 247, 383

\bibitem[Szeifert et al. (1993)]{szeifert93} Szeifert, Th., Stahl, O., Wolf, B. et al., 1993, A\&A, 280, 508

\bibitem[Tautvaisiene et al. (2007)]{taut07} Tautvaisiene, G., Geisler, D., Wallerstein, G. et al., 2007, AJ, 134, 2318

\bibitem[Tolstoy et al. (1995)]{tolstoy95} Tolstoy, E., Saha, A., Hoessel, J.G., McQuade, K., 1995, \aj, 110, 1640

\bibitem[Udalski et al. (2001)]{udalski01} Udalski, A., Wyrzykowski, L., Pietrzy\'nski, G, 2001, AcA 51, 221

\bibitem[Vink (2009)]{vink09} Vink, J.S., 2009, astro-ph 0905.3338

 \end{thebibliography}

\end{document}